\documentclass{emulateapj}

\usepackage{amsmath}
\usepackage{graphicx}
\usepackage{natbib}
\usepackage{bm}

\usepackage{epsfig}

\shorttitle{Evolution of shocks and turbulence in major cluster mergers}
\shortauthors{Paul et al.}

\begin{document}

\title{Evolution of shocks and turbulence in major cluster mergers}

\author{S. Paul\altaffilmark{1,4}, L. Iapichino\altaffilmark{2}, F. Miniati\altaffilmark{3}, J. Bagchi\altaffilmark{4} and K. Mannheim\altaffilmark{1}}
\altaffiltext{1}{Institut f\"ur Theoretische Physik und Astrophysik, Universit\"at W\"urzburg, Am Hubland, D-97074 W\"urzburg, Germany}
\altaffiltext{2}{Zentrum f\"ur Astronomie der Universit\"at Heidelberg, Institut f\"ur Theoretische Astrophysik, Albert-Ueberle-Strasse 2, \\ D-69120 Heidelberg, Germany}
\altaffiltext{3}{Physics Department, Wolfgang-Pauli-Strasse 27, ETH-Z\"urich, CH-8093 Z\"urich, Switzerland}
\altaffiltext{4}{The Inter-University Center for Astronomy and Astrophysics, Pune University Campus, Pune 411 007, India}

\begin{abstract}
We performed a set of cosmological simulations of major mergers in galaxy clusters, in order to study the evolution of merger shocks and the subsequent injection of turbulence in the post-shock region and in the intra-cluster medium (ICM). The computations have been performed with the grid-based, adaptive mesh refinement (AMR) hydrodynamical code Enzo, using a refinement criterion especially designed for refining turbulent flows in the vicinity of shocks. When a major merger event occurs, a substantial amount of turbulence energy is injected in the ICM of the newly formed cluster. 
Our simulations show that the shock launched after a major merger develops an ellipsoidal shape and gets broken by the interaction with the filamentary cosmic web around the merging cluster. 
The size of the post-shock region along the direction of shock propagation is of the order of $300\ \mathrm{kpc}\ h^{-1}$, and the turbulent velocity dispersion in this region is larger than $100\ \mathrm{km\ s^{-1}}$.
We performed a scaling analysis of the turbulence energy within our cluster sample. The best fit for the scaling of the turbulence energy with the cluster mass is consistent with $M^{5/3}$, which is also the scaling law for the thermal energy in the self-similar cluster model. This clearly indicates the close relation between virialization and injection of turbulence in the cluster evolution. As for the turbulence in the cluster core, we found that within $2\ \mathrm{Gyr}$ after the major merger (the timescale for the shock propagation in the ICM), the ratio of the turbulent to total pressure is larger than 10\%, and after about $4\ \mathrm{Gyr}$ it is still larger than $5 \%$, a typical  value for nearly relaxed clusters. Turbulence at the cluster center is thus sustained for several Gigayears, which is substantially longer than typically assumed in the turbulent re-acceleration models, invoked for explaining the statistics of observed radio halos. Striking similarities in the morphology and other physical parameters between our simulations and the `symmetrical radio relics' found at the periphery of the merging cluster Abell~3376 are finally discussed. In particular, the interaction between the merger shock and the filaments surrounding the cluster could explain the presence of `notch-like' features at the edges of the double relics.
\end{abstract}


\section{Introduction}
\label{intro}

Galaxy groups and clusters are the largest virialized objects to have arisen in the process of cosmic structure formation. These structures have been forming by assembling through hierarchical clustering of matter driven by gravity. The large scale distribution of matter in the Universe has a web-like structure consisting of an interconnected network of large filaments, voids, and sheets, where clusters form at the nodes of this matter distribution \citep{Bond_1996_Natur,Einasto_1997_A&AS,Doroshkevich_1996_MNRAS}. How such complex structures  arise out of the primeval density fluctuations has been an enduring quest. The cosmic shock waves which are produced by accretion and halo mergers play an important role in the process of hierarchical structure formation, yet their origin and
evolution is not very well understood \citep{Bykov_SSRV_2008}.

Previous numerical simulations \citep{Miniati2_ApJ_2000,Miniati_2002_MNRAS,Ryu_ApJ_2003,Pfrommer_MNRAS_2006,Skillman_2008_arxiv,Vazza_2009_MNRAS} make a distinction between two classes of shocks, namely the `internal' and `external' shocks. The former group indicates weaker (Mach number $\mathcal{M} < 5$) shocks found in the hot ICM resulting from merging events, whereas the latter ones are generated by the in-fall of cold, unprocessed baryons on accreting structures, thus producing shocks with stronger temperature gradients and consequently larger Mach numbers ($\mathcal{M} > 10$). Both external and internal shocks dissipates a large fraction of kinetic energy in the ICM. Strong collisionless astrophysical shocks are notably capable of producing high energy cosmic-ray particles (CR) via diffusive shock acceleration mechanism (\citealt{Blandford_1987_PhsRep}, for a review). These shocks thus have been proposed as acceleration sites of the CR particles emitting the observed non-thermal radio emission in clusters \citep{Ensslin_AA_1998,Roettiger_ApJ_1999,Miniati_ApJ_2001,Miniati_2003_MNRAS}. The cosmic $\gamma$-ray background may also be generated by the same electron populations through inverse-Compton scattering of the Cosmic Microwave Background Radiation \citep{Loeb_2000_Natur, Miniati_2002_MNRAS, Bagchi_Sci_2006, Miniati_2007_ApJ}.

Major cluster mergers, where the mass ratio of the infalling halos
approaches unity, are among the most energetic
events of the Universe (\citealt{Ricker_2001_ApJ}; the energy release
from a merger of two clusters of mass $10^{15}\ M_{\sun}$, moving with
$\sim 1000\ \mathrm{km\ s^{-1}}$ relative to each other, may approach
$10^{64}\ \mathrm{erg}$).

Merger shocks are particularly interesting for the injection of
volume-filling turbulence in the ICM
\citep{Subramanian_2006_MNRAS}. If a turbulent flow is established in
the ICM, turbulent dissipation acts on a significantly longer
timescale than shocks.  Turbulence can in principle re-accelerate the
ambient electrons and amplify magnetic fields by the dynamo action.

In the last few decades, a substantial amount of observational
evidences for major merger events in clusters have been collected,
mainly in X-rays as well as in optical observations, and only recently
in radio wavelengths. The radio structures found in galaxy clusters
are mainly of two types, the central radio halos and the peripheral
relics \citep{Ferrari_2008_SSRv.}. Among the peripheral radio relics,
a special type of `symmetrical radio structure' has been found (also
known as `Radio Gischts', cf.~\citealt{Kempner_2004_rcfg}). These
structures are mainly found in cluster of galaxies
where supposedly a massive merger is
ongoing. They are thought to be the tracers of nearly symmetric
shocks propagating through the ICM, emanating
out of the merging center.
Such a structure was seen for the first time in the
cluster A3667, reported by \citet{Rottgering_MNRAS_1997}. A
spectacular double-radio arc which could be part of a gigantic (size
of the order of $2\ \mathrm{Mpc}$) ring-like formation was discovered
in the cluster Abell 3376 (A3376) by \citet{Bagchi_Sci_2006}. Apart from these
two, only recently a few more examples of such structures have been observed; the
filamentary merging cluster ZwCl 2341.1+000 which shows both diffuse radio  emission 
along the filaments \citep{Bagchi_NA_2002,Giovannini_A&A_2010} and peripheral double relics separated by about $2.2\ \mathrm{Mpc}$ \citep{Weeren_2009_A&A}
and the double radio relics in the clusters A2345 and A1240,
reported by \citet{Bonafede_2009_A&A}.

Concerning the CR acceleration occurring in such shocks, two processes
are important in the present context, namely the first-order Fermi acceleration
at shock front \citep{Drury_1983a_RepProgPh,Blandford_1987_PhsRep}
and, perhaps, the second-order Fermi acceleration in the turbulent
post-shock region~\citep{Schlickeiser_1989_ApJ}.

Due to the inherent complexity of the merger events, hydrodynamical simulations have been the main investigation tool for the study of hierarchical buildup of cosmic structures. Furthermore, idealized halo merger setups have often been used, in order to simplify the physical description of mergers, and to explore the parameter space in a rather controlled way \citep{Heinz_2003_MNRAS,Asai_2004_ApJ,Takizawa_2005_ApJ,Takizawa_2005_ASR,Asai_2005_ASR,Asai_2007_ApJ,Xiang_2007_MNRAS}. Clearly, hydrodynamical simulations of cluster evolution in a cosmological framework are somehow complementary to the above cited approach, but provide the final test-bed for the idealized studies, and the necessary link to the observations. The increase of the availability of computational resources and of `simulation catalogs' make cosmological simulations more suitable for a thorough study of merger events.

In this work we have simulated several major merger events in a
non-artificial setup with the hydrodynamical code Enzo
\citep{O'Shea_2005_Springer}. The aim is mainly to account for the
effect of the propagation of merger shocks and their role in injecting
turbulence in the ICM, in particular in the post-shock
region. Emphasis will be therefore put on quantitative and
morphological features of merger shocks and the subsequently generated
turbulence in ICM. In a second part we also analyze the turbulent flow in the cluster core, a region which is better probed by observations, and where the radio halos are considered manifestations of the merging activity. Both for the turbulence in the post-shock region and in the cluster core, the scaling properties with the cluster mass have been studied. 

In the present work, we do not explore a detailed
connection between simulations and cluster radio observations,
although we will be presenting some tentative comparison of our
simulated turbulent shock structures with the interesting Mpc-scale
radio structures observed in merging clusters (specifically A3376), revealing appealing close similarities. In a future publication
we wish to present a more detailed comparative analysis of the shock
induced synchrotron radio structures simulated numerically, with the
giant peripheral radio structures actually observed in dynamically
active clusters.

The work is structured as follows: details of the performed simulations and the criteria for selecting mergers are described in Section~\ref{simuD}. The analysis of the performed runs and the related results are presented in Section~\ref{results}. The robustness of these results with respect to numerical convergence is shown in Section~\ref{resol}, and finally the results are discussed, compared with observations and summarized in Section~\ref{disc}.

\section{Numerical simulations}
\label{simuD}

\subsection{Setup of the simulations}
\label{setup}

We performed a series of cosmological simulations of cluster mergers using the Adaptive Mesh Refinement (AMR), grid-based hybrid (N-body plus hydrodynamical) code Enzo v.~1.0 \citep{O'Shea_2005_Springer}. Our simulations assume a flat $\Lambda$CDM background cosmology with parameters $\Omega_{\mathrm \Lambda} = 0.7$, $\Omega_{\mathrm m} = 0.3$, $\Omega_{\mathrm b} = 0.04$, $h = 0.7$, $\sigma_8 = 0.9$, and $n = 1$. The simulations have been initialized at redshift $z = 60$ using the \citet{Einstein_ApJ_1999} transfer function, and evolved to $z = 0$. An ideal equation of state was used for the gas, with $\gamma = 5/3$. Cooling physics and feedback are neglected, because they play only a minor role in the merger problem  that we investigate in this paper. Furthermore, neglecting
cooling  processes imply that the simulated virialized objects obey self-similar scaling laws \citep{Bryan_1998_ApJ}. This turns out   useful as it allows us to extrapolate results obtained for the modest mass simulated objects ($M \simeq 10^{14}\ M_{\sun}$) to the more
  massive and actually observed structures.

The simulation box has a comoving size of $128\ \mathrm{Mpc}\ h^{-1}$. It is resolved with a root grid (AMR level $l = 0$) of $64^3$ cells and $64^3$ N-body particles. The mass of each particle in this grid is $8.3 \times 10^{11}\ M_{\sun}$. A static grid ($l = 1$) is nested inside the root grid. It has a size of $64\ \mathrm{Mpc}\ h^{-1}$ and is resolved in $64^3$ cells and $64^3$ particles (particle mass $1.03 \times 10^{11}\ M_{\sun}$). Inside this grid, in a volume with side of $32\ \mathrm{Mpc}\ h^{-1}$, a further nested grid at $l=2$ is added ($64^3$ cells and $64^3$ particles, with mass $1.3 \times 10^{10}\ M_{\sun}$), and grid refinement from level $l = 3$ to $l = 6$ is enabled. The linear refinement factor $N$ is set to 2, allowing an effective spatial resolution of $31.25\ \mathrm{kpc}\ h^{-1}$ at the maximum refinement level. The dependence of our problem on resolution issues is discussed in Section~\ref{resol}.
The two static grids and the region where AMR is allowed are nested on the location of a cluster merger, identified according to the prescriptions discussed in Section~\ref{merging-setup}.

The mesh refinement is a particularly critical point of the setup of such merger simulations. As will be shown in the following, a shock is launched soon after a merger event, and it propagates from the cluster center to the outskirts up to the virial radius and beyond, through a medium with gradually decreasing density. An AMR criterion based only on baryon and dark matter (DM) over-density could not track the evolution of this transient in a satisfactory way. Therefore, for a better resolution of the post-shock regions, we coupled the AMR on over-density with a refinement criterion suitable for refining the development of turbulence in the vicinity of propagating shocks. In particular, since shocks are associated with flows with negative divergence of the velocity field, we used an AMR criterion based on the local variability of the rate of compression of the flow (the negative time derivative of the divergence $d = \mathbf{\nabla \cdot v}$). This criterion was developed by \citet{Schmidt_2008_AA} and used in the AMR simulations of \citet{Iapichino_MNRAS1_2008} and \citet{Iapichino_2008_MNRAS}. According to it, a cell is considered for refining if the local value of the compression rate $c(\mathbf{x},t)$ fulfills the criterion
 \begin{equation}
\label{local}
c(\mathbf{x},t) \ge \langle c \rangle_i(t) + \alpha \lambda_i(t)
\end{equation}
where $\lambda_i$ is the maximum between the average $\langle c
\rangle $ and the standard deviation of $c$ in the grid patch $i$, and $\alpha$
is a tunable parameter, set to 5.0 for balancing resolution and computational efficiency.

In the over-density criterion, a cell is refined if
\begin{equation}
\label{threshold}
\rho_\mathrm{i} > f_\mathrm{i} \rho_0 \Omega_\mathrm{i} N^l\,\,,
\end{equation}
where $\rho_0 = 3 H_0^2 / 8 \pi G$ is the critical density. The formula holds for both baryons and DM.
In this work the parameters for over-density are set to $f_{\mathrm{b}} = f_{\mathrm{DM}} = 4.0$ (cf.~\citealt{Iapichino_2008_MNRAS}).

\subsection{Cluster mergers}
\label{merging-setup}

A preliminary, DM-only run was performed in order to identify DM halos, and to select merger events to be re-simulated with the setup described in the previous section. The halos have been identified using the HOP algorithm \citep{Eisenstein_1998_ApJ}.

We focus our study on merger events which occurred at $z < 0.7$ and between halos with mass $M > 10^{13}\ M_{\sun}$ 
at the time of merger. As a criterion for selecting major mergers, we rejected events with a mass ratio between the merging clumps $\Delta_{\mathrm m} < 0.5$, a value taken 
as the minimum possible ratio at which the cores of the two merging halos get destroyed to form a new cluster core \citep{Salvador_1998_ApJ}. Also events starting at $z < 0.25$ were excluded, because the merger shock did not have enough time to propagate through the ICM before $z = 0$ in those cases. 

After checking from the mass history that a major merger was ongoing, we visually inspected the data for prominent merging shocks; in this way, seven representative mergers were chosen in our DM-only simulation. Clusters are then re-simulated as in Section \ref{setup}. Since the computational box has periodic boundary conditions, the same volume is shifted and the nested grids are centered for each run on different locations, corresponding to different forming clusters. The selected events span different mass ratios, merger redshifts, and total cluster masses.
The features of the simulated mergers are summarized in Table \ref{table_cluster}, where the runs are identified by letters from $A$ to $H$. Run $H$ is a special comparison case representing a relaxed cluster, which did not experience any massive merger in its recent evolution to $z = 0$ (cf.~Figure~\ref{rlx_actv_turb}$a$).

\begin{deluxetable}{cccccc}
\tablewidth{0pt}  
\tablecaption{Overview of the performed simulations}
\tablehead{
\colhead{Run}           &
\colhead{Redshift of}     &
\colhead{$M_1$} &
\colhead{$M_2$} &
\colhead{$\Delta_{\mathrm m}$} &
\colhead{$M_{\mathrm f}$} \\
\colhead{}           &
\colhead {the merger}      &
\colhead{$[10^{13}\ M_{\sun}]$ } &
\colhead{$[10^{13}\ M_{\sun}]$}          &
\colhead{}  &
\colhead{$[10^{14}\ M_{\sun}]$}
}
\startdata
$A$   & 0.7   & 2.38  & 1.65  & 0.69  & 1.08 \\
$B$   & 0.7   & 9.75  & 5.95  & 0.61  & 2.21 \\
$C$   & 0.5   & 5.51  & 3.29  & 0.60  & 1.69 \\
$D$   & 0.5   & 2.98  & 2.84  & 0.95  & 0.88 \\
$E$   & 0.4   & 9.86  & 5.63  & 0.57  & 4.73 \\
$F$   & 0.3   & 7.23  & 4.23  & 0.59  & 2.62 \\
$G$   & 0.25  & 2.11  & 1.05  & 0.50  & 0.59 \\
$H$   &\nodata&\nodata&\nodata&\nodata& 4.80 \\
\enddata
\tablecomments{The masses of the merging clumps (third and fourth column) are computed at the merging time (redshift indicated in the second column). The fifth column contains the corresponding mass ratio for the clumps, whilst in the sixth column the final mass of the cluster at $z = 0$ is reported.}
\label{table_cluster}
\end{deluxetable}

\section{Results}
\label{results}

\subsection{Morphological evolution of the merger shock}
\label{morphology}

As already stated, our attention will be initially focused on the propagation of merger shocks, and on the evolution of turbulence in the downstream region. It is instructive to introduce those results with a description of the main features of the major mergers. Among all the simulated merging clusters in Table~\ref{table_cluster}, the merger $F$ has the best morphology with fully developed and nicely resolved shock structures, which even propagate beyond the virial radius before fading away. This case was therefore chosen as the reference run for our analysis.

\begin{figure*}
\epsscale{1}
\plotone{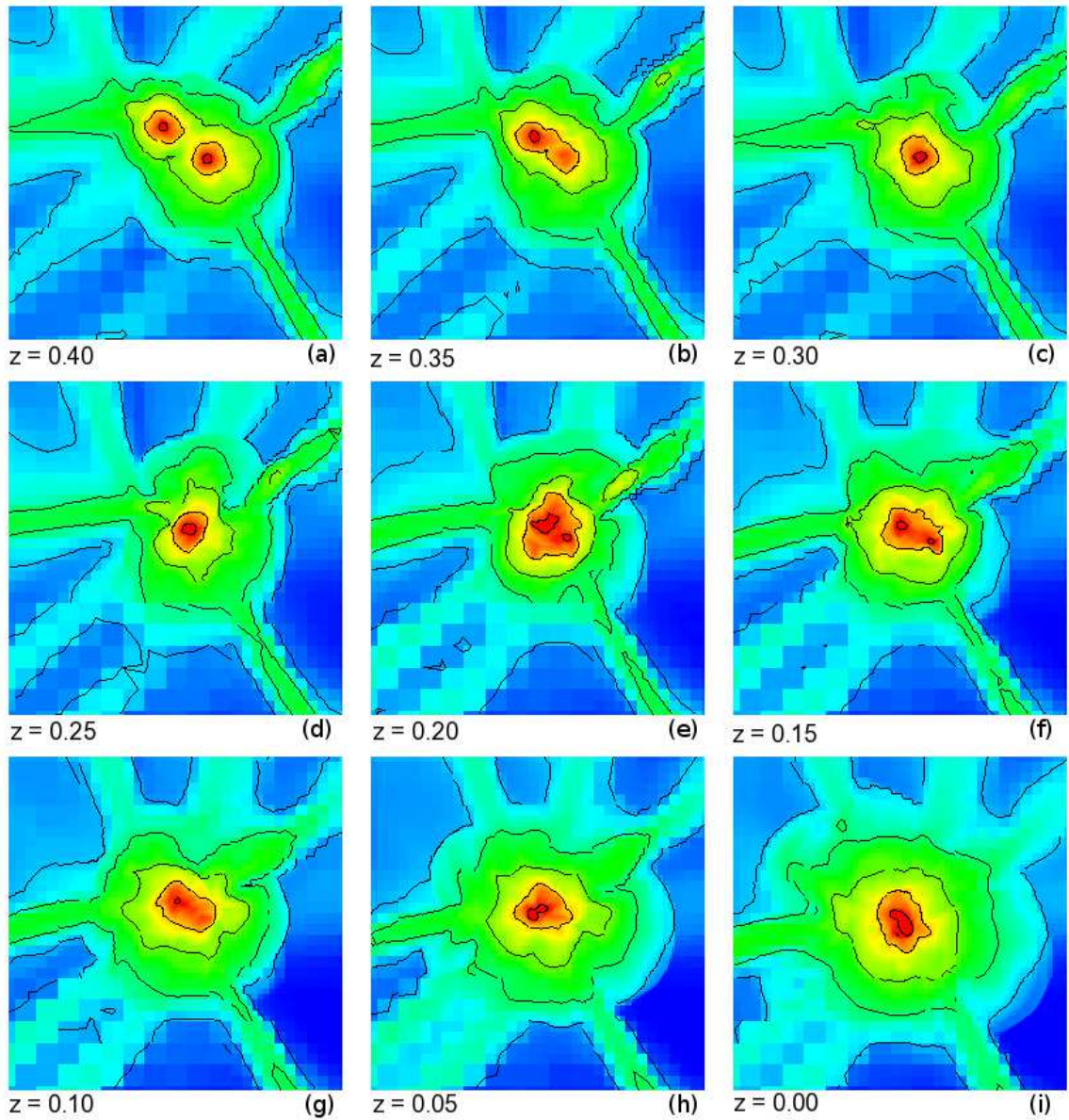}\\
\caption{The evolution of merger $F$ is shown in density slices. The redshift is indicated at the lower left of each panel, and a identification letter is at the lower right. Each panel has a size of $7.7 \times 7.7\ \mathrm{Mpc}\ h^{-1}$ and is parallel to the $yz$ plane. The baryon density is color coded, and also represented by contours.}
\label{density}
\end{figure*}

The morphological evolution of merger $F$ is shown in Figure~\ref{density}. Two sub-clumps approach each other 
with a relative velocity of $980 \ \mathrm{km\ s^{-1}}$,  
collided for the first time at $z \simeq 0.3$ (Figure~\ref{density}$c$) and then pass through a core oscillation phase. The merging cores are still distinctly  visible at least until $z = 0.1$ (Figure~\ref{density}$g$) before the final coalescence at $z = 0.05$. The web-like network of filaments is also clearly visible around the forming structure.

\begin{figure*}
\epsscale{1}
\plotone{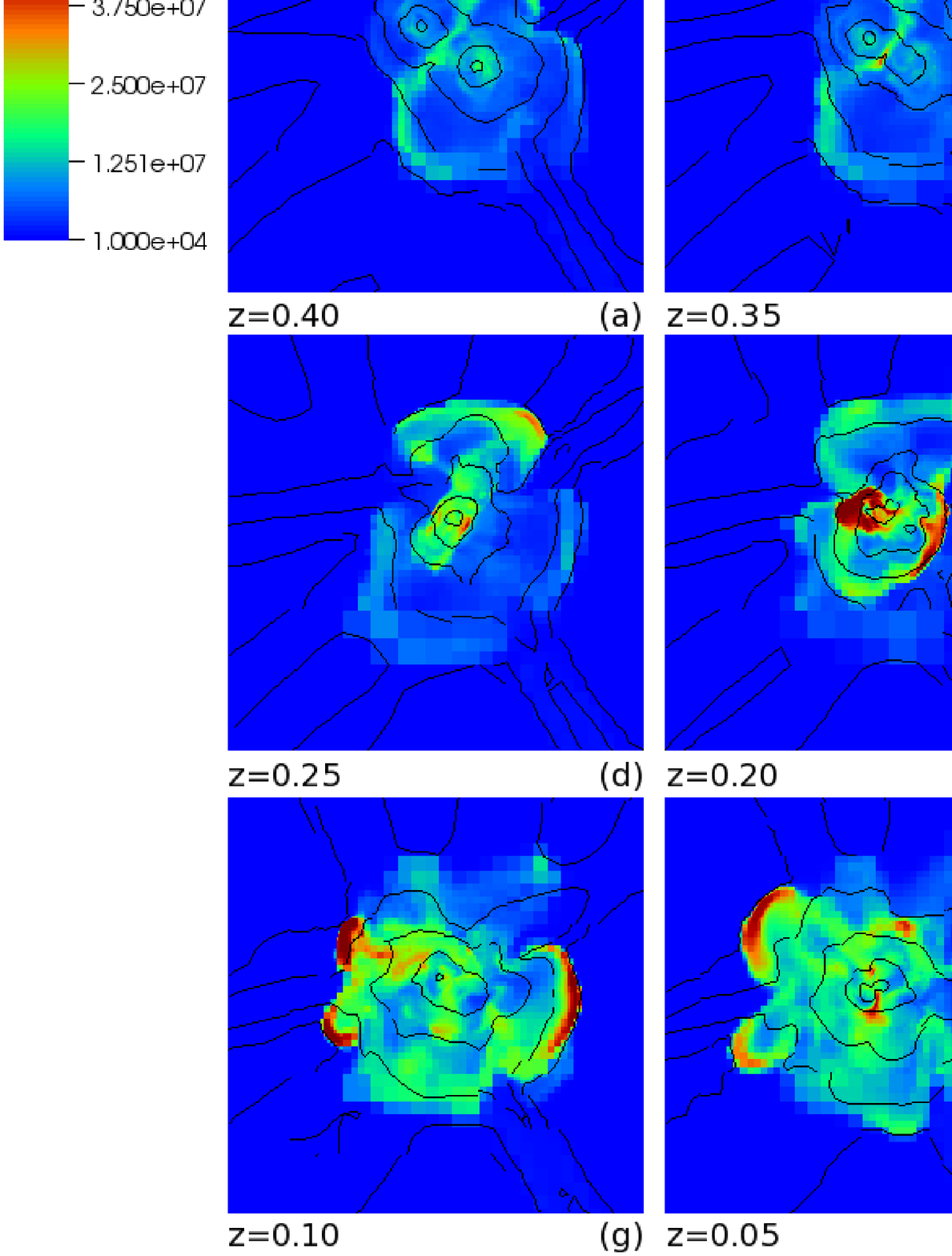} \\
\caption{Same as Figure~\ref{density}, but temperature is here color coded, with density contours overlayed.
In order to display the lower temperature structures the upper limit of the temperature (color scale) has been cutoff at $5.0 \times 10^7\ \mathrm{K}$ (Deep brown-red). The maximum temperature at the central region of the cluster goes upto $ 10^8\ \mathrm{K}$ in certain cases and temperature behind the shock goes beyond $5.0 \times 10^7\ \mathrm{K}$.}\label{temperature}
\end{figure*}

The most prominent effect during a cluster merger is the evolution of the baryonic component, whose energy budget is significantly altered by the event. The gas in the ICM is severely attracted in the forming potential well, eventually generating a shock wave which propagates through the intra-cluster gas of the newly formed cluster. Part of the kinetic and gravitational energy of the merger event is thus dissipated into the ICM.

This crucial feature of mergers is better followed by the evolution of temperature (Figure~\ref{temperature}). The temperature increase is first driven by compression at the center of the forming cluster (panels \ref{temperature}$a$ to \ref{temperature}$c$), and subsequently the shock is launched and propagates outwards. It can be followed in the simulation even after it covers few virial radii from the cluster center (at $z = 0$, $R_{\mathrm{vir}} = 0.92\ \mathrm{Mpc}\ h^{-1}$ for the cluster in run $F$). The maximum temperature in the central region of the cluster exceeds $10^8\ \mathrm{K}$ at $z = 0.2$, several times larger than the cluster virial temperature (about $3 \times 10^7\ \mathrm{K}$).

\begin{figure*}
\epsscale{1}
\plotone{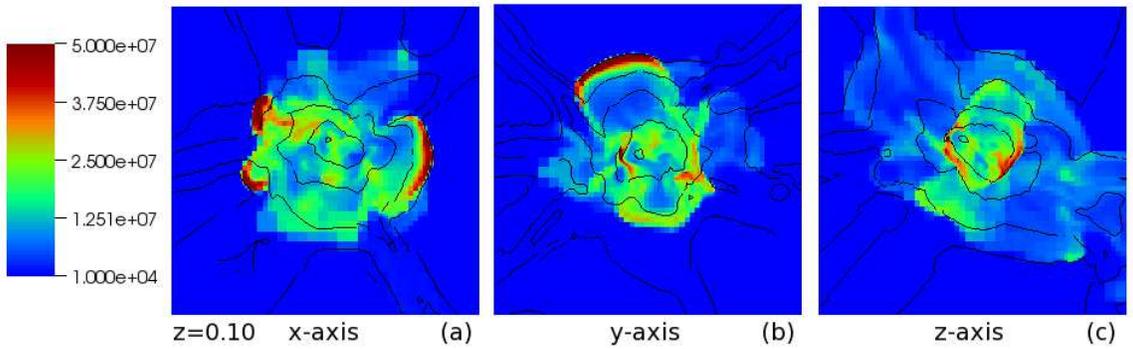} \\
\caption{Evolution of the shock in temperature, as seen from slices in three different planes of the computational volume. The panels refer to $z = 0.1$ and to slices perpendicular to the $x$-axis (left column), to the $y$-axis (central column) and to the $z$-axis (right column), respectively. Each panel has a size of $7.7 \times 7.7\ \mathrm{Mpc}\ h^{-1}$ and is cut along the center of mass of the system.}
\label{shock-shape}
\end{figure*}

The shape of the emerging shock depends on the mass of the merging clumps and on the geometry of the merger. In the case of merger $F$, 
the shock front has a roughly ellipsoidal 
shape, with the arcs more pronounced along the merger axis. The morphology of the evolving shock front from different lines of sight is shown in Figure~\ref{shock-shape}.  
The propagation velocity of the shock 
is initially up to $1500\ \mathrm{km\ s^{-1}}$ and only slightly decreases in time with the expansion of the shock.
We computed the Mach number of the two most prominent shock arcs, to the left and to the right with respect to the cluster center,  
by applying the Rankine-Hugoniot jump conditions. Between $z = 0.25$ and 0.15, $\mathcal{M}$ is in the range between 2.5 to 7. The larger values are reached for the arc on the right-hand side of the panels in Figure~\ref{temperature}, resulting in a mild asymmetry.

As this shock propagates out of the ICM of the newly formed cluster, it interacts with the surrounding filaments. The interaction with the web-like cosmic structure causes the breaking of the merger shock in separate sections, as clearly visible by comparing the bottom row of Figure~\ref{temperature} and Figure~\ref{density}. 
This interesting feature is obviously not modeled so far in simulations of idealized mergers (e.g.~\citealt{Ricker_2001_ApJ,Ritchie_MNRAS_2002,Mitchell_2009_MNRAS}), where symmetric bow-like shocks propagates unimpeded outwards.

\subsection{Generation and evolution of turbulence past the merger shock}
\label{turbu}

The development and propagation of a large scale shock is a remarkable consequence of the hierarchical growth of halos undergoing major mergers. Besides the already described effects for the morphology, the energetics and the heating of the ICM, in this section we discuss the injection and evolution of turbulence in the post-shock region.

Velocity fluctuations are a distinctive feature of turbulent flows. It is therefore straightforward to relate the generation of turbulence with the production of vorticity occurring at curved shock, expressed by the curl of the Euler's equation \citep{Landau_1959_book}:

\begin{equation}
\label{vort}
\frac{\partial {\bm \omega}}{\partial t} = {\bf \nabla \times} ({\bf v \times {\bm \omega}}) - \frac{{\bf \nabla} p \; {\bf \times \nabla} \rho}{\rho^2}
\end{equation}
where the second term at the right-hand side is non-vanishing in curved shocks.

\begin{figure*}
\epsscale{1}
\plotone{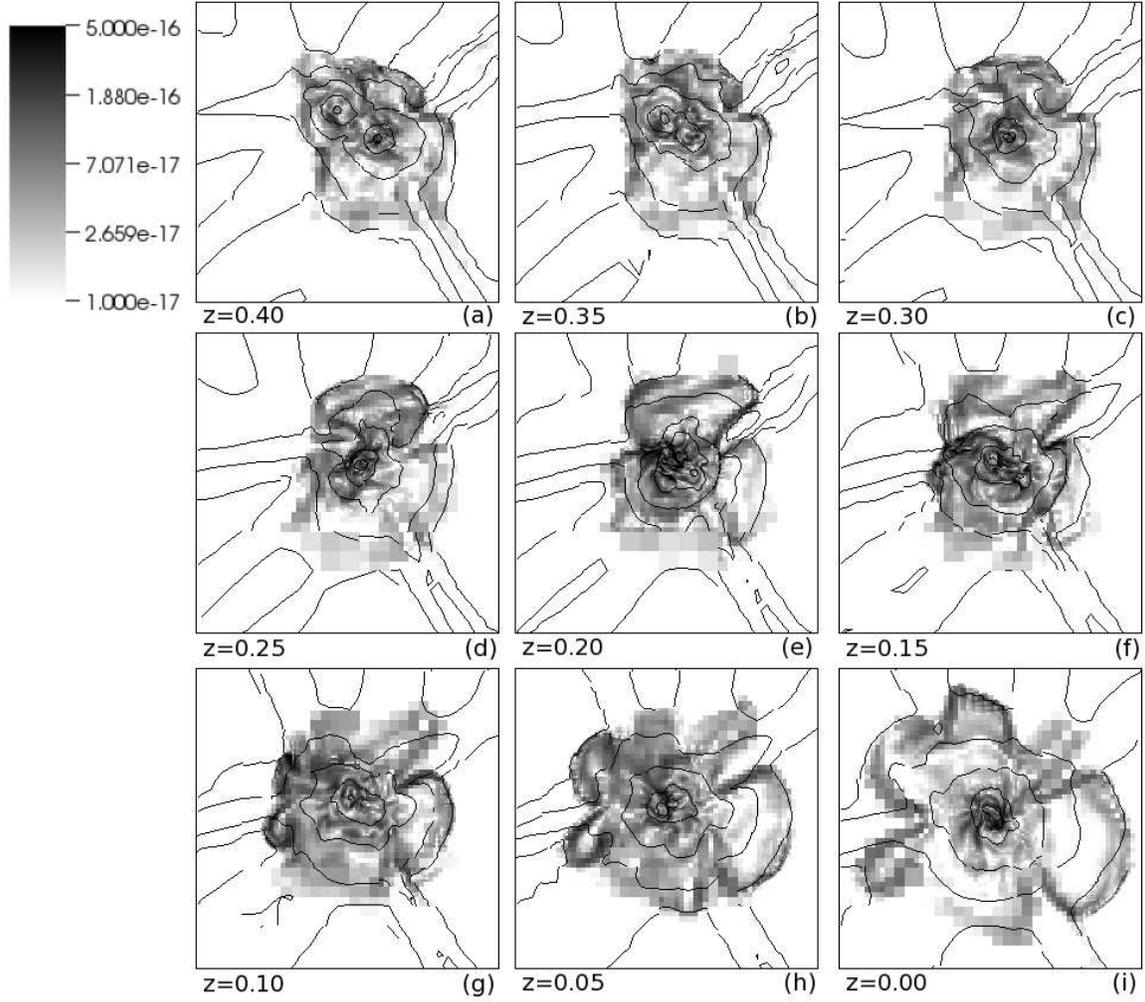}
\caption{Same as Figure~\ref{density}, but slices of the norm of the vorticity ${\bm \omega} = {\bf \nabla \times v}$ are shown, with density contours overlayed. The vorticity has dimension $[t^{-1}]$, and is reported in $\mathrm{s^{-1}}$ unit in the gray-scale at the upper left corner.}
\label{vorticity}
\end{figure*}

This theoretical expectation is confirmed by our simulations. We computed the vorticity from the velocity field, using a four-point symmetric method for computing the spatial derivatives, with fourth-order space accuracy. In the reference run $F$, vorticity is produced after the merger, just behind the shock, and propagates along with it. This process is shown in Figure~\ref{vorticity}: some level of vorticity is associated to both clumps before merging (Figure~\ref{vorticity}$a$) and also to the center of the newly formed cluster (Figure~\ref{vorticity}$i$), but the magnitude of ${\bm \omega}$ tracks markedly the generation of turbulence in the post-shock region (for $z < 0.2$, Figures~\ref{vorticity}$e$ to~\ref{vorticity}$i$). Different from minor mergers \citep{Subramanian_2006_MNRAS,Iapichino_2008_MNRAS,Maier_2009_ApJ}, in the case under consideration a simple visual inspection of Figure~\ref{vorticity} suggest that major mergers stir the ICM effectively, resulting in a very volume-filling production of turbulence. In Figure~\ref{vorticity} we also observe that values $\omega \gtrsim 5 \times 10^{-17}\ \mathrm{s^{-1}}$ correspond, according to the definitions of \citet{Kang_ApJ_2007}, to a vorticity parameter (representing the number of local eddy turnovers) $\tau \gtrsim 10$, namely to a relatively large vorticity and to a full developed turbulence. In agreement with \citet{Kang_ApJ_2007}, such high values are reached in the ICM after the merger, as well as in the post-shock region.

\begin{figure}
\epsscale{1}
\plotone{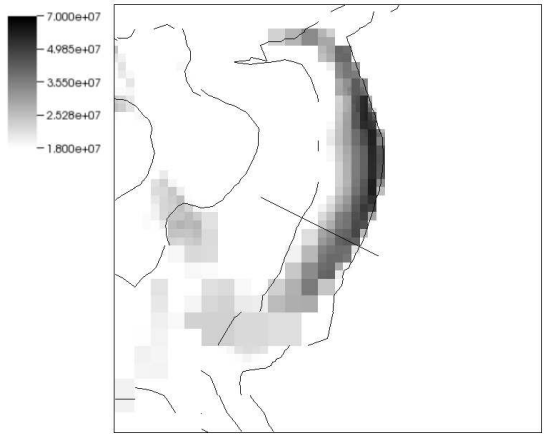}
\caption{Temperature slice at $z = 0.1$ (cf.~Figure~\ref{temperature}$g$), with a size on a side of $3.2\ \mathrm{Mpc}\ h^{-1}$. Density contours are superimposed. The short line crossing the post-shock region is used for computing the one-dimensional profile shown in Figure~\ref{FWHM}.}
\label{vorti_ext}
\end{figure}

\begin{figure}
\epsscale{1}
\plotone{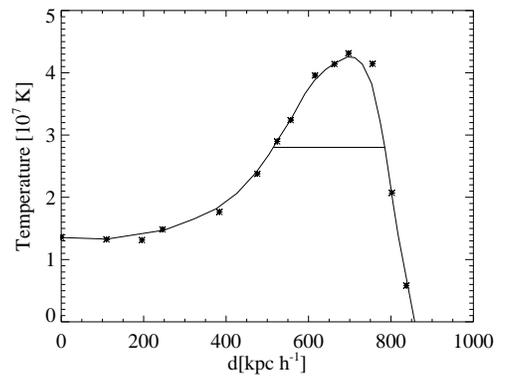}
\caption{Profile of temperature along the line shown in the slice of Figure~\ref{vorti_ext}. The interpolating line is a $\beta$-spline fitting to the data points indicated with stars. The horizontal solid line indicates the level at $T = T_{\mathrm{max}} - \Delta T / 2$ (with $\Delta T = T_{\mathrm{max}} - T_{\mathrm{min}}$), with the lowest temperature past the shock propagation $T_{\mathrm{min}} \simeq 1.3 \times 10^7\ \mathrm{K}$, and the meaning of the symbols described in the text.}
\label{FWHM}
\end{figure}

The size of the post-shock region along the direction of propagation of the shock can be roughly estimated from Figure~\ref{vorti_ext}. In Figure~\ref{FWHM} this issue is addressed in a more quantitative way, with the analysis of a one-dimensional temperature profile computed along a line crossing the shock, drawn in Figure~\ref{vorti_ext}. Along this line the temperature rises abruptly from the pre-shock to the post-shock region (from right to left along the $x$-axis in Figure~\ref{FWHM}) and then it decreases more gently towards the cluster center. As an estimate for the size of the post-shock region, we computed its width at half maximum along the line used for the profile in Figure~\ref{FWHM}: given $T_{\mathrm{max}} \simeq 4.3 \times 10^7\ \mathrm{K}$ and $T_{\mathrm{min}} \simeq 1.3 \times 10^7\ \mathrm{K}$, the length of the region in Figure~\ref{FWHM} where $T > T_{\mathrm{max}} - \Delta T / 2$ (with $\Delta T = T_{\mathrm{max}} - T_{\mathrm{min}}$) is about $270\ \mathrm{kpc}\ h^{-1}$. A similar result can be obtained from the analysis of the vorticity. 
This size can be taken as a typical order-of-magnitude estimate for regions past propagating shocks in major merger events.

\begin{figure}
\epsscale{1}
\plotone{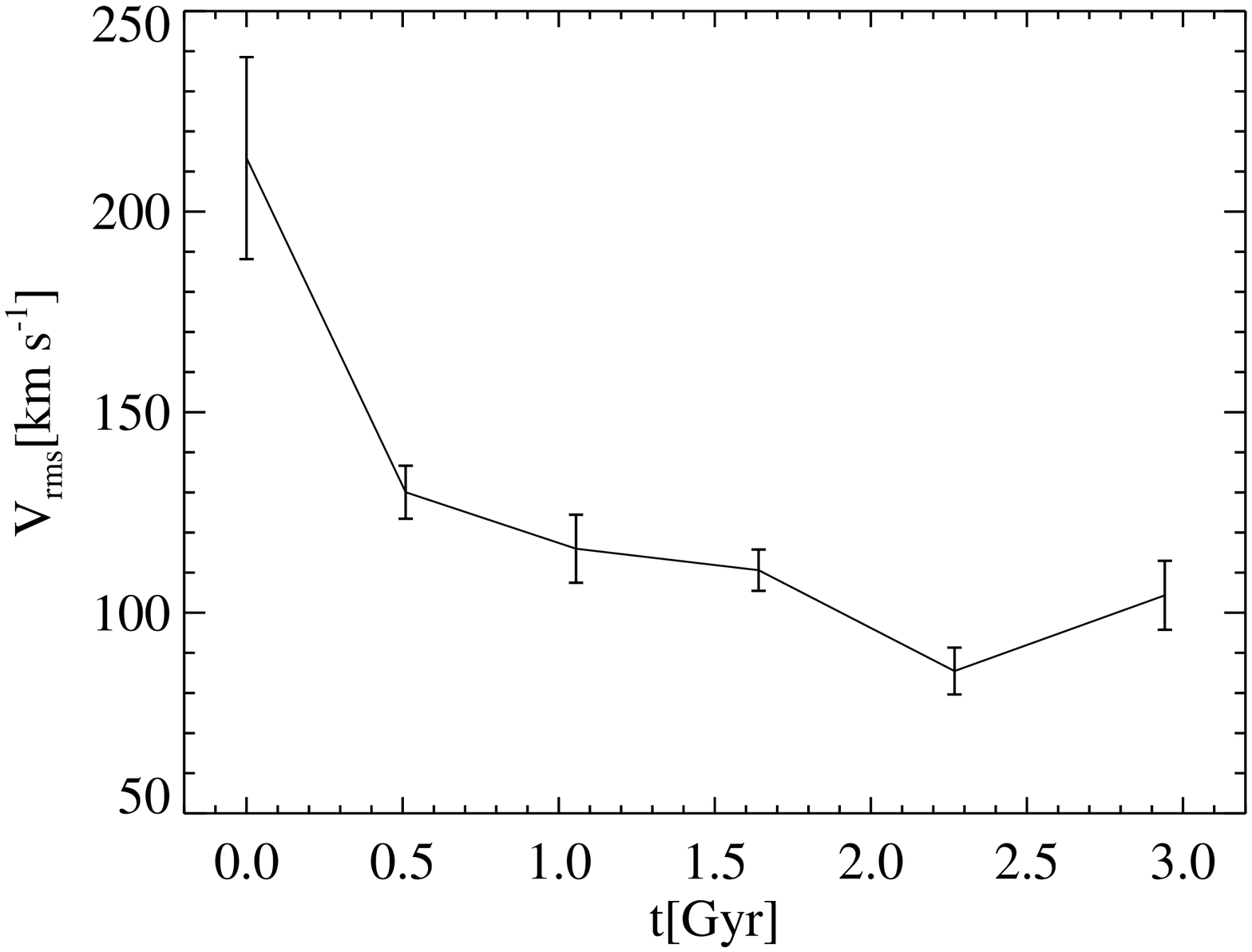}
\caption{Temporal evolution of $v_{\mathrm {rms}}$ in the post-shock region in cluster $F$. The error bars represent the standard errors of the computation of $v_{\mathrm {rms}}$.}
\label{vrms_time}
\end{figure}

Above in this section we used the vorticity as an indicative quantity of the turbulent state of the flow. A further diagnostic of the magnitude of the turbulent motions which is often used is the root mean square (henceforth rms) velocity, defined on a generic domain as

\begin{equation}
\label{vrms}
v_{\mathrm {rms}} =  \sqrt{ \frac{\sum_{i} m_{i}(v_{i} - \langle v \rangle )^{2}} {\sum_{i} m_{i}}}
\end{equation}

where the sum is performed on the analysis domain, and $m_{i}$ and $v_i$ are the mass contained and the velocity magnitude in the cell $i$, respectively. The quantity $\langle v \rangle$ is the average value of $v_i$ in the analysis volume; from an operational point of view, the calculation of $v_{\mathrm {rms}}$ involves the definition of this average reference velocity, in order to distinguish between the bulk velocity flow and the velocity fluctuations (cf.~\citealt{Dolag_MNRAS_2005}). In our case, $\langle v \rangle$ and $v_{\mathrm {rms}}$ have been computed on a sphere with a diameter of $256\ \mathrm{kpc}\ h^{-1}$, placed inside the post-shock region. The size of this analysis region has been chosen to be similar to that of the post-shock region, in order to resolve the latter with the largest possible number of cells. The shock arc shown in Figure~\ref{vorti_ext} was chosen for the analysis, and, to partly avoid spurious fluctuations of $v_{\mathrm {rms}}$, we averaged the results of many (from three to more than ten, depending on the evolving size of the region) analysis spheres and computed the standard error to show its statistical reliability. The method is then applied to every output of the simulation, and the results are plotted in  Figure~\ref{vrms_time}. The time evolution of $v_{\mathrm {rms}}$ shows that the velocity dispersion is larger when the shock has just emerged, and then it decays with a timescale of about $1.5\ \mathrm{Gyr}$.

\begin{figure}
\epsscale{1}
\plotone{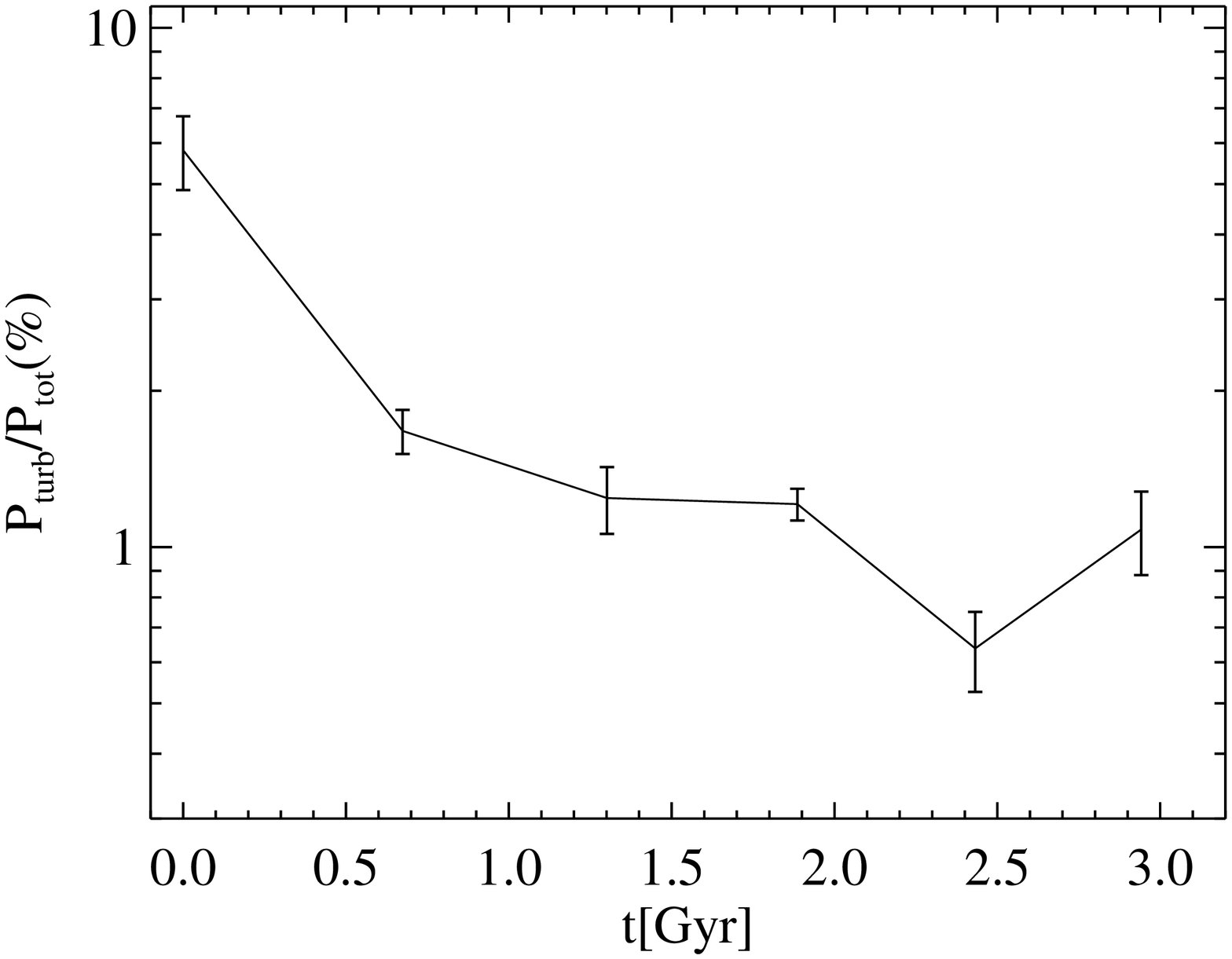}
  \caption{As in Figure~\ref{vrms_time}, but the ratio of turbulent to total pressure is here reported.}
\label{press_ratio}
\end{figure}

We make use of the velocity dispersion to define the turbulent pressure support as the ratio of the turbulent pressure ($P_{\mathrm{turb}}$) to the total pressure ($P_{\mathrm{tot}}$):

\begin{equation}
\label{support}
\frac{P_{\mathrm{turb}}}{P_{\mathrm{tot}}} =  \frac{v_{\mathrm{rms}}^{2}/3}{ k T/ (\mu m_{\mathrm p})+ v_{\mathrm {rms}}^{2}/3}
\end{equation}
where $k$ is the Boltzmann constant, $\mu = 0.6$ is the mean molecular weight in a.m.u. and $m_{\mathrm{p}}$ is the proton mass. The turbulent to total pressure ratio (Figure~\ref{press_ratio}) peaks at relatively large values, close to $10\ \%$, and then it decays in a way similar to $v_{\mathrm {rms}}$. In both temporal evolutions, at about $2.9\ \mathrm{Gyr}$ from the shock emergence, one can notice a mild increase of the turbulence diagnostics, caused by the interaction of the merger shock with the reached accretion shock. 

\begin{figure}
\plotone{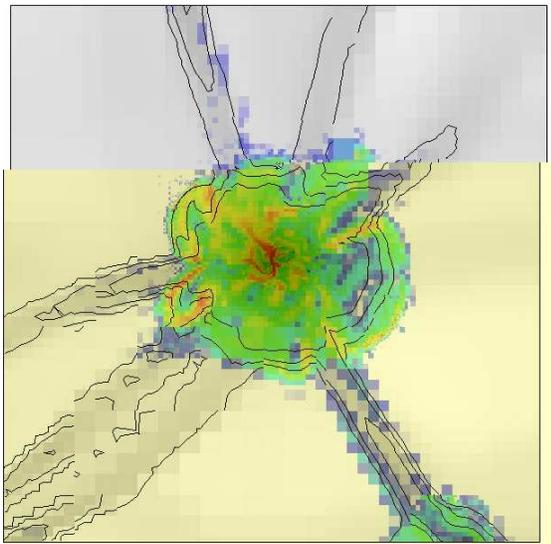}
\caption{Slice of $12.8 \times 12.8\ \mathrm{Mpc}\ h^{-1}$, showing the squared norm of the vorticity $\omega^2$ for the cluster $F$ at $z = 0.05$. The vorticity squared is color coded, whereas the density is represented in shaded gray-scale and contours, in order to better highlight the filaments surrounding the cluster.}
\label{filvort}
\end{figure}

The shock propagation does not stir the ICM only in the post-shock region. An interesting by-product, highlighted in Figure~\ref{filvort}, is the interaction between merger shocks and filaments. It is already known that the warm-hot baryons flowing along the filaments results in a turbulent flow, when it mixes with the cluster gas \citep{Nagai_APJ_2003,Maier_2009_ApJ}. Here, as an additional effect, we can see that a substantial level of vorticity (comparable to that injected in the post-shock region) is generated in the regions past the merger shock and surrounding filaments. In Figure~\ref{filvort} such `collars' can be clearly observed for the second and third filament starting from twelve-o'clock clockwise, and among the two filaments on the left-hand side. The turbulence injection in these zones is probably related to the shearing (Kelvin-Helmholtz) instability between the shocked gas moving outwards and the filament gas flowing inwards. The level of vorticity is much larger than the turbulence associated with the baroclinic generation at the filament accretion shock, in regions far from the clusters (well visible for the third filament, in the lowest part of Figure~\ref{filvort}). In Section~\ref{disc} we further discuss about the link of this feature and radio observations of double relics.

\subsection{Comparison with a relaxed cluster: turbulence injection in the ICM}
\label{relaxed}

For a better understanding we compare the properties of the merger in run $E$ with those of a relaxed cluster ($H$, in Table~\ref{table_cluster}). This merger is different from the run $F$ used in the previous analysis and it was chosen primarily because it has nearly the same final mass as the cluster $H$ at $z = 0$.

\begin{figure*}
\epsscale{1}
\plotone{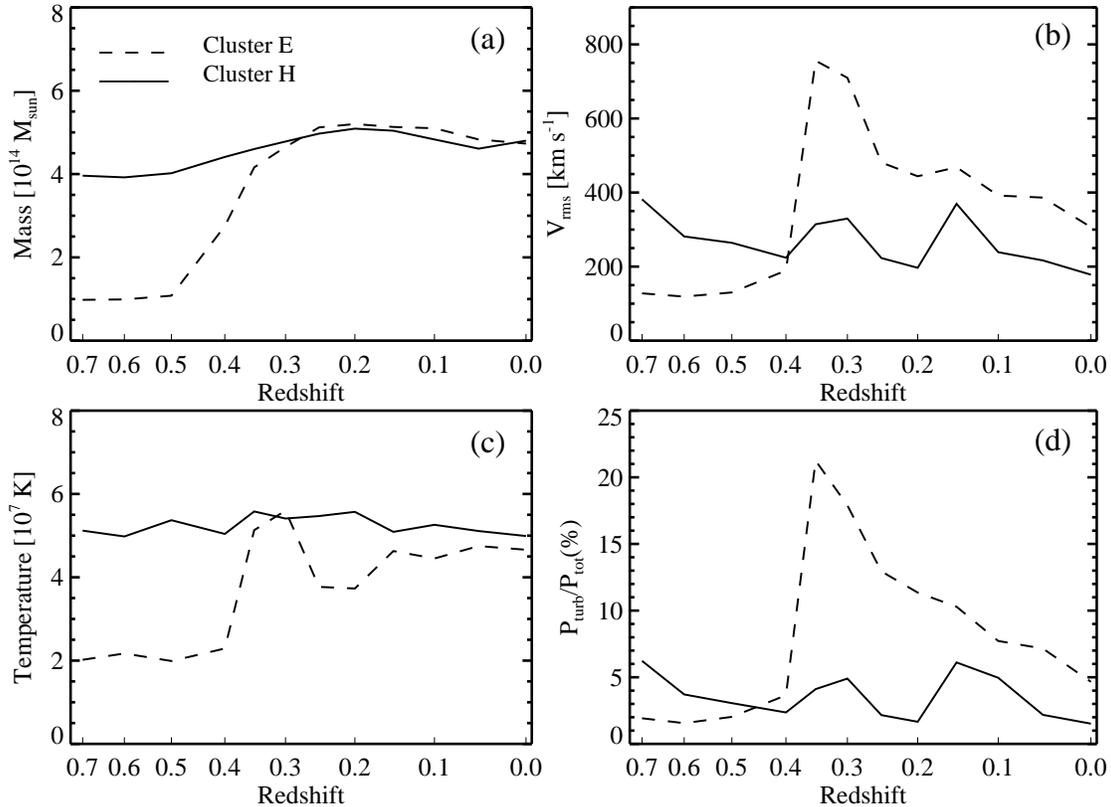}
\caption{Comparison of cluster $E$ (dashed lines) and cluster $H$ (solid lines), through the temporal evolution of selected quantities. With the exception of panel $a$, the computation is based on mass-weighted averages computed on an analysis sphere with the diameter of $512\ \mathrm{kpc}\ h^{-1}$, centered at the cluster center. The $x$ axis shows time, indicated by the corresponding redshift, and spans for $6\ \mathrm{Gyr}$. Panel $a$: evolution of the cluster mass. Panel $b$: rms velocity. Panel $c$: temperature. Panel $d$: ratio of turbulent and total pressure.}
\label{rlx_actv_turb}
\end{figure*}

For both clusters we performed an analysis of mass-weighted averages of selected quantities (Figure~\ref{rlx_actv_turb}), computed inside a spherical volume of diameter $512\ \mathrm{kpc}\ h^{-1}$ placed at the cluster center. The centers are defined as the locations of the peak of DM density, as provided by the HOP tool.

The different accretion histories of the two clusters are clearly distinguishable in Figure~\ref{rlx_actv_turb}$a$, where the temporal evolution of the cluster mass is reported. A gradual increase in mass can be noticed for the cluster $H$, whereas cluster $E$ has an abrupt mass gain starting at $z = 0.5$, consistent with a major merger scenario. The minor fluctuation in the evolution of mass are due to the uncertainties of the analysis tool.

The accretion events in the cluster histories are even more apparent in the evolution of $v_{\mathrm{rms}}$ (Figure~\ref{rlx_actv_turb}$b$). This quantity has a steep increase in run $E$ during the major merger, reaching an average value of about $800\ \mathrm{km\ s^{-1}}$, and then decreases steadily, halving its maximum value in about $2\ \mathrm{Gyr}$ but still remaining larger than in the relaxed cluster, until $z = 0$. There is some delay between the mass increase and the rise of the velocity dispersion, probably linked to the emergence of the merger shock. In the relaxed cluster, the overall trend of $v_{\mathrm{rms}}$ is more regular, although there are two moderate peaks which are very likely related with minor mergers.

The comparison of the temperature evolution (Figure~\ref{rlx_actv_turb}$c$) is particularly instructive about the role of turbulence in dissipating the kinetic energy injected by mergers. The two clusters have nearly the same final mass and, as expected from simple scaling and virial arguments, end up
at $z = 0$ with a similar average temperature $T$. Whilst for the relaxed cluster $T$ shows almost no evolution during the last $6\ \mathrm{Gyr}$, because at $z = 0.7$ the structure is already virializing and relaxing, for the merging cluster $E$ the temperature increase is slowing down only near $z = 0$. The peak of $T$ at $z \simeq 0.3$ is connected with the propagation of the merger shock through the analysis sphere; this event stirs the ICM (Figure~\ref{rlx_actv_turb}$b$), and the dissipation of the turbulent motions is evidently correlated with the increase of internal energy. The temperature increase is not limited to the peak at $z = 0.3$ but extends to later times, with a trend reminiscent of the double shock model for mergers presented by \citet{McCarthy_2007_MNRAS}.

Finally in Figure~\ref{rlx_actv_turb}$d$ the turbulent pressure support $P_{\mathrm{turb}}/P_{\mathrm{tot}}$ (equation~\ref{support}) is shown. The turbulent contribution to the total pressure during the major merger is significant, being larger than $10\%$ for about $2\ \mathrm{Gyr}$ after the merger.

From Figure~\ref{rlx_actv_turb}$d$ we infer that the level of turbulent pressure support (which is equivalent to the corresponding ratio between energies) typical for minor merger events in otherwise relaxed clusters is of the order of $5 \%$ at most, comparable to the value found in the cluster core by \citet{Iapichino_2008_MNRAS}. For simplicity, hereafter we assume that $5 \%$ of turbulent pressure support is the threshold between relaxed and perturbed cluster cores. In case of cluster $E$, the whole decay time to $P_{\mathrm{turb}}/P_{\mathrm{tot}} = 5\%$ is of the order of $4\ \mathrm{Gyr}$.

\begin{figure}
\epsscale{1}
\plotone{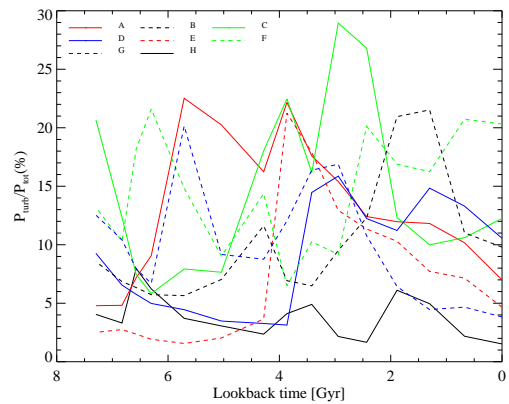}
\caption{Time evolution of pressure fraction of turbulence in all studied clusters of Table~\ref{table_cluster}. The solid black line is used for the relaxed cluster $H$, rest are the merging clusters as indicated in the figure legends.}
\label{rlx_actv_turb_all}
\end{figure}

When all clusters of our sample are analyzed using this criterion, it turns out that the run $E$ is a rather typical example of turbulent decay after a cluster merger. For few clusters, the support level remains even larger than $10 \%$ for most of the evolution after the major merger (see Figure~\ref{rlx_actv_turb_all}), probably because of later repeated events. The shortest decay time, found in one cluster of our sample, is $1\ \mathrm{Gyr}$. The relatively large values for the decay timescale for turbulent motions fit well with similar estimates in idealized cluster mergers (e.g., \citealt{Ricker_2001_ApJ}).

\subsection{Scaling relations for turbulence}
\label{scaling}

The scaling of ICM features with the cluster mass is a useful and informative tool to study the physics of galaxy clusters. Self-similar models of clusters (since the seminal work of \citealt{Kaiser_1986_MNRAS}) rely on the assumption that gravity is the  main driver of the evolution of such structures, whereas departures from simple scaling laws imply some more complicate physics at work (for example, in the case of entropy in cluster cores; see \citealt{Borgani_2008_SSR} for a review).

The scaling of turbulence energy in clusters has been investigated with a semi-analytic approach by \citet{Cassano_2005_MNRAS}, and by \citet{Vazza_2006_MNRAS} using SPH simulations. In both cases a scaling of the turbulence energy with the cluster mass $E_{\mathrm{turb}} \propto M^\alpha$, with $\alpha \simeq 5/3$, as expected for the thermal energy in the virial case. This is consistent with the assumption, invoked in the model of \citet{Cassano_2005_MNRAS}, that the turbulence energy injected in the clusters is proportional to the internal energy of the forming structure, which in turn depends on its mass.

In this section we present a tentative scaling study, performed on our simulation set. Obviously, our relatively small number of runs hardly covering one order of magnitude in cluster mass will result in  rather large uncertainties on the slope of the fit. But nevertheless it is interesting to compare the above cited results with those from our sample, consisting only of clusters with ongoing major mergers. Firstly the turbulence injected in the cluster core by major mergers is investigated, and then we will focus on the post-shock region of merger shocks.

\begin{figure}
\epsscale{1}
\plotone{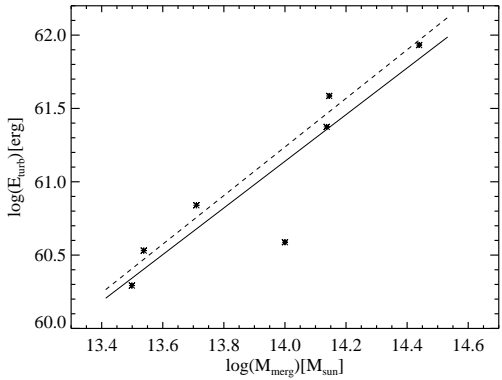}
\caption{The turbulence energy $E_{\mathrm{turb}}$ (defined in the text) is here plotted against the merger mass $M_{\mathrm{merg}}$, for the merging clusters of our sample (Table~\ref{table_cluster}). The solid line is the best fit to the data points, corresponding to a scaling law $E_{\mathrm{turb}} \propto M^{1.6}$, whereas the dashed line (with a slope of 1.66) is the best fit computed by excluding the cluster $F$ (the point at $\log M = 14.0$).}
\label{merger-scaling}
\end{figure}

We define the merger mass $M_{\mathrm{merg}}$ as the cluster virial mass (including DM) computed with the utility {\tt enzo\_anyl}, assuming an over-density $\delta = 200$, and the turbulence energy $E_{\mathrm{turb}} = 0.5\ M_{\mathrm{sph}}\ v_{\mathrm{rms}}^2$, where $v_{\mathrm{rms}}$ is the velocity dispersion in an analysis sphere with the diameter of $512\ \mathrm{kpc}\ h^{-1}$ centered at the cluster center, and $M_{\mathrm{sph}}$ is the baryon mass contained in this sphere. Both the centers of the analysis sphere and that one for the virial mass computation are defined from the peak in DM density. 
The virial radii $R_{200}$ in the cluster sample range between $0.43$ and $0.95\ \mathrm{Mpc}\ h^{-1}$ right after the shock emerged, with a moderate growth at later times, so that the analysis sphere has a size that is always well contained within the cluster volume.

In Figure~\ref{merger-scaling} the scaling properties of these quantities are studied. The calculation for each cluster has been performed at the time of the merger, as reported in Table~\ref{table_cluster}. 
The best-fit to the scaling exponent $\alpha$ provides a value of $1.6 \pm 0.3$. The uncertainty is large, but nonetheless the result is in agreement with the previous studies of \citet{Cassano_2005_MNRAS} and \citet{Vazza_2006_MNRAS}.

\begin{figure}
\epsscale{1}
\plotone{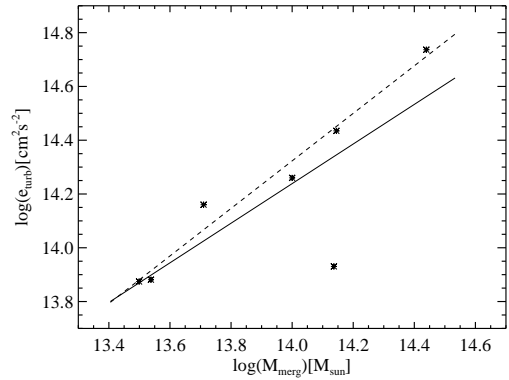}
\caption{Same as Figure~\ref{merger-scaling}, but the effective turbulence energy in the post-shock region $e_{\mathrm{turb}}$ is on the $y$-axis. The solid line is the best fit to the scaling law, with exponent $\alpha = 0.74$, and the dashed line (obtained by excluding the outlier point of cluster $C$) has $\alpha = 0.88$.}
\label{post-scaling}
\end{figure}

We are not aware of any previous investigation of scaling laws for the turbulence injected in the post-shock region by merger shocks. In Figure~\ref{post-scaling} our results on this property are presented; the specific turbulence energy $e_{\mathrm{turb}}$ is defined as $0.5 \times v_{\mathrm{rms}}^2$, where the velocity dispersion is computed in the post-shock region as described in Section~\ref{turbu}. 
Here the fitting scaling law has a slope of $0.74 \pm 0.25$. 

A calculation of the mass in the post-shock region has been avoided, because of the related excessive uncertainties. Assuming that $M_{\mathrm{post-shock}} \propto M_{\mathrm{merg}}$, and defining the total turbulence energy as $E_{\mathrm{turb}} = M_{\mathrm{merg}}\ e_{\mathrm{turb}}$, we retrieve also in this case a scaling law with a dependence compatible with $M_{\mathrm{merg}}^{5/3}$, thus indicating also in this case a close link between gravitational infall, virialization, and injection of turbulence energy.

It is immediate	to see that both the cluster samples in Figure~\ref{post-scaling} and, to a lesser extent, in Figure~\ref{merger-scaling} contain an outlier. The outlier in Figure~\ref{merger-scaling}, with $M_{\mathrm{merg}} \simeq 10^{14} M_{\odot}$, is the cluster $F$, while in Figure~\ref{post-scaling} ($\log(M_{\mathrm{merg}}/M_{\odot}) \simeq 14.4$) it is cluster $C$. We verified by inspecting the data that cluster $C$ underwent an off-axis major merger with a relatively large impact parameter, 
leading to a phase of core rotation before the final coalescence. It can be speculated that, due to this evolution, the energy transferred to the merger shock has been smaller than in the other cases, or released on longer timescales (cf.~\citealt{Mitchell_2009_MNRAS}). As for cluster $F$, a phase of core oscillation has been also observed (Section~\ref{morphology}). This feature can potentially affect the analysis also from a numerical viewpoint, since it makes difficult for the analysis tool to univocally detect the cluster center, introducing inaccuracies in the computation of the virial mass and of the turbulence energy.

Removing the outlier from the best-fit calculation provide a scaling exponent of $1.66 \pm 0.13$ (without cluster $F$) in Figure~\ref{merger-scaling} and $0.88 \pm 0.07$ (without cluster $C$) in Figure~\ref{post-scaling}, respectively.

\section{Resolution study}
\label{resol}

The choice of a refinement criterion and of its relevant thresholds and parameters is a very delicate task in AMR simulations. Cosmological simulations of galaxy clusters are favored by the clumped behavior of these objects: criteria based on baryon or dark matter over-density, when properly set, are able to catch the relevant structures with a good compromise between accuracy and saving of computational resources \citep{O'Shea_2005_ApJS,Heitmann_2008_CSD}.

Unfortunately the simulations presented in this work are extremely challenging from this point of view, because we are mainly interested in the evolution of the shocked region, whose size at late times is comparable with the cluster size, and that for our purpose should be carefully refined. Numerical tests lead to the design of the setup presented in Section~\ref{setup}. In this setup the final effective resolution is not particularly high, if compared with similar grid-based cluster simulations. 
We verified that the adopted resolution level is adequate for refining the post-shock region in major mergers, as the numerical tests presented below will show.

A key feature of grid-based cosmological simulation is the force resolution in the DM section of the code. We recall that in the second static nested grid at $l = 2$, $64^3$ N-body particles are allocated in a volume with size of $32\ \mathrm{Mpc}\ h^{-1}$, resulting in a particle mass of $1.3 \times 10^{10}\ M_{\sun}$ for our standard setup. In order to study the convergence properties of our computations, test simulations were performed with $32^3$ and $128^3$ N-body particles, both in the root grid and in the two static grids. Both tests gave unsatisfactory results: the latter was computationally not feasible because of constraints on the available memory, whilst the former was so poorly resolved that the two clumps did not even show a merger shock. For this reason, we had to reduce the explored range of the resolution study to two further calculations, with root grid resolution from $l = 0$ to $l = 2$ of $48^3$ and $96^3$ grid cells and N-body particles. In the following 
 these runs will be labeled as $F48$, and $F96$, respectively. Both runs allow for six additional AMR levels, so for both the effective resolution changes by a factor 1.5 with respect to the reference run $F$.

\begin{figure*}
\epsscale{1}
\plotone{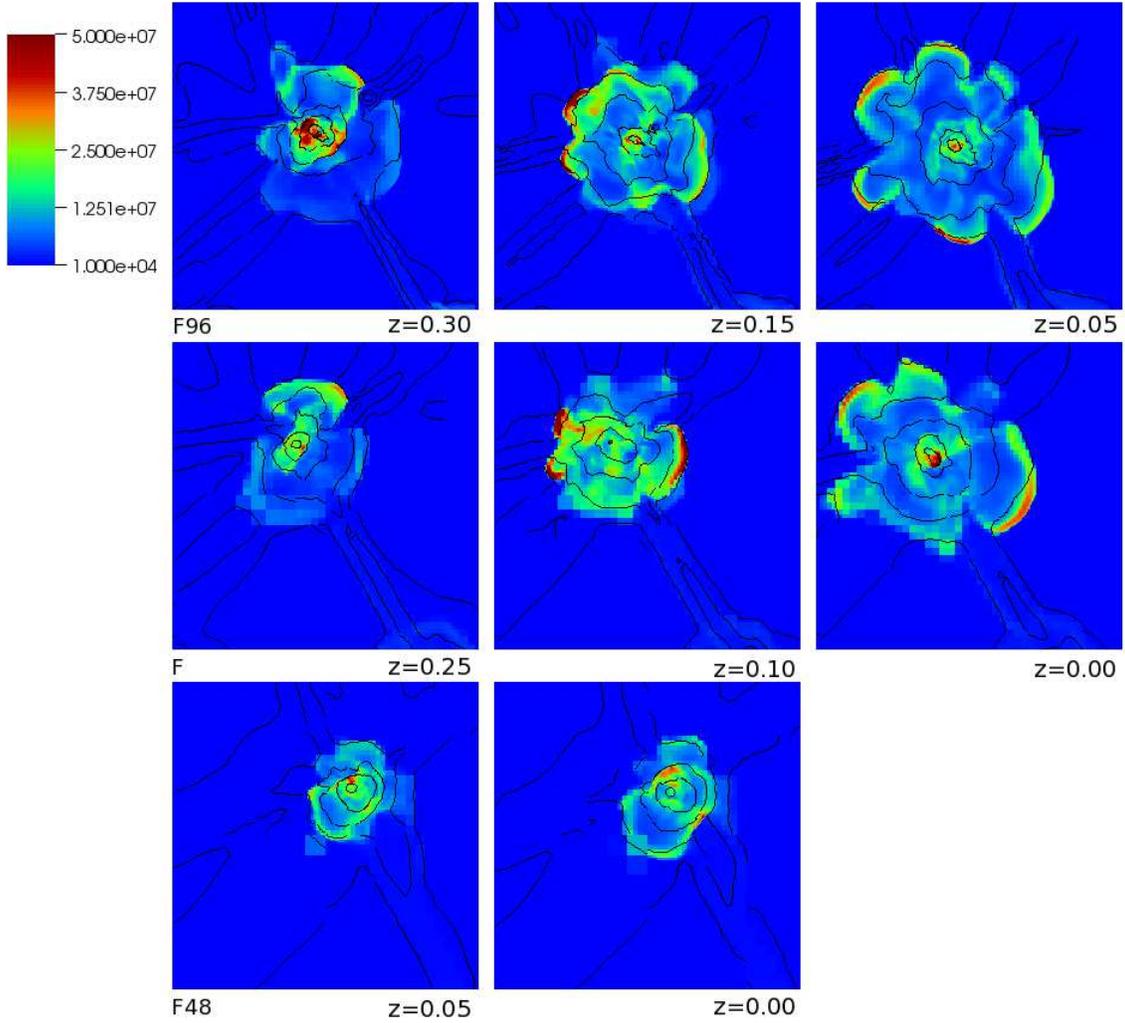}
\caption{Temperature slices with density contours overlayed, similar to Figure~\ref{temperature}, but each having a size of $10.2 \times 10.2\ \mathrm{Mpc}\ h^{-1}$. The evolution of runs $F96$, $F$, and $F48$ is shown in the upper, middle, and lower panels, respectively. The redshifts in the vertical rows do not correspond along the different simulations, because of the time delay effect discussed in the text.}
\label{temp_dens_all}
\end{figure*}

One of the most interesting morphological differences is a significant time delay in the emergence of the merger shock for different resolutions. In particular, the shock is launched at $z \simeq 0.3$ for run $F96$, at $z \simeq 0.25$ for run $F$, and at $z \simeq 0.05$ for run $F48$. In this last case, the evolution is therefore followed only for a very short time before $z = 0$. The reason for this time delay probably lies in the transient nature of merger shocks, which appears to depend critically on the morphology of the merging substructures at different resolutions. In order to permit a better morphological comparison, the time delay has been compensated in the panels of Figure~\ref{temp_dens_all}. A qualitative analysis of the temperature slices shows that there is a good morphological match between runs $F$ and $F96$, whereas run $F48$ cannot be judged on the same grounds, because of its excessive delay.

\begin{figure}
\epsscale{1}
\plotone{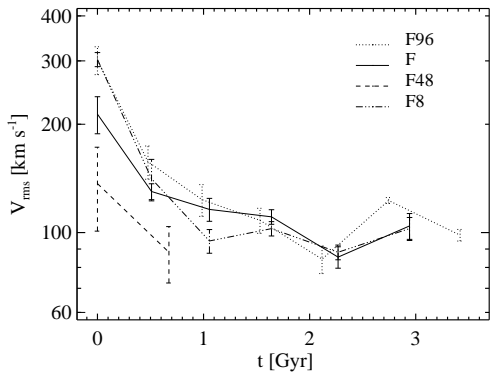}
\caption{The temporal evolution of the velocity dispersion $v_{\mathrm{rms}}$ in the post shock region is computed, using the method described in Section~\ref{turbu}, for the runs $F96$, $F$, $F48$ and $F8$ (see legend). The different evolutions have been shifted in time, in order to approximately compensate for the time delay discussed in the text. The initial points have a large intrinsic uncertainty, due to the inability of placing an adequate number of examining spheres when the post-shock region has a very limited size.}
\label{vrms_all}
\end{figure}

A more quantitative comparison between the simulations with different root grid resolution is provided by the analysis of the velocity dispersion $v_{\mathrm{rms}}$ in the post-shock region. We applied the procedure described in Section~\ref{turbu} for the calculation of $v_{\mathrm{rms}}$ using a number of analysis spheres downstream of the merger shock, and we repeated for the three runs under examination. Figure~\ref{vrms_all} shows a moderate convergence of the values of $v_{\mathrm{rms}}$ between the runs $F96$ and $F$; run $F48$ apparently underestimates the velocity dispersion, although an accurate comparison is hindered by the too short shock evolution. The size of the post-shock region (Figure~\ref{FWHM}) is also converged in runs $F96$ and $F$, whereas in $F48$ the shock has not emerged enough at $z = 0$ to permit the comparison.

From this resolution study we conclude that runs $F96$ and $F$ give results which are consistent each other, whereas $F48$ under resolves the physical processes in the ICM which are relevant in this work. Though $F96$ is still marginally manageable within our available computational resources, we decided to work on a root grid resolution of $64^3$, because it turned out to be more suitable for running a whole series of simulations, as those performed in this study.

Besides spatial resolution, obviously the AMR grid resolution plays a relevant role in properly modeling the vorticity production. In our numerical scheme, vorticity is not explicitly computed by a separate equation, but is derived by the velocity field. Is our maximum AMR level sufficient for resolving the latter in a sufficient way? 

In order to address this point, we repeated the run $F$ with a maximum AMR level $l = 8$, instead of $6$; this run will be labeled $F8$. The effective resolution in the run $F8$ reaches $7.8\ \mathrm{kpc}\ h^{-1}$, making this simulation equivalent to the computations presented in \citet{Iapichino_2008_MNRAS} and \citet{Maier_2009_ApJ}. 

\begin{figure}
\epsscale{1}
\plotone{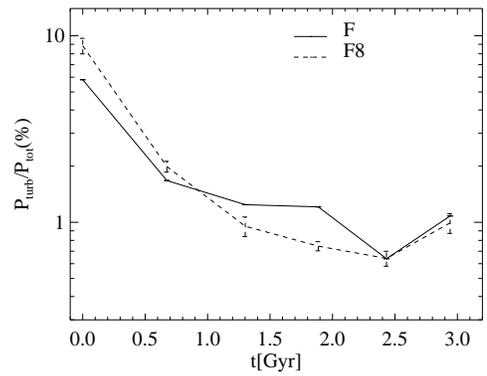}
\caption{Time evolution of the pressure ratio (equation~\ref{support}) in the post-shock region, for the runs $F$ (solid line) and $F8$ (dashed line).}
\label{Post_AMR}
\end{figure}

\begin{figure}
\epsscale{1}
\plotone{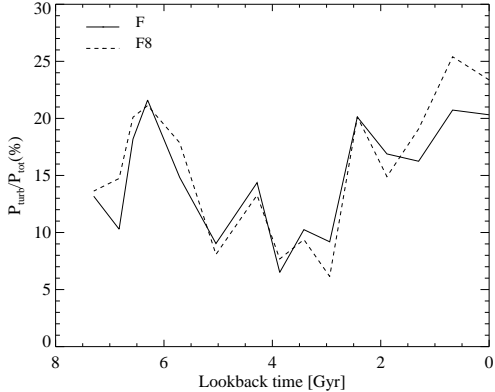}
\caption{Time evolution of the pressure ratio (equation~\ref{support}) in the cluster core for the simulations $F$ (solid line) and $F8$ (dashed line).}
\label{Central_8AMR}
\end{figure}

After having verified that most of the post-shock region benefits of the improved resolution at $l = 7$ and $8$, we computed there the evolution of $v_{\mathrm{rms}}$, shown in Figure~\ref{vrms_all}. In this case, the velocity evolution for run $F8$ resembles both runs $F96$ and $F$, with a maximum discrepancy (about $50\%$) right after the shock is launched, where it is most difficult to perform the calculation of $v_{\mathrm{rms}}$. No significant time delay has been found in the shock evolution, with respect to run $F$. Interestingly, the pressure ratio $P_{\mathrm{turb}}/P_{\mathrm{tot}}$ in the post-shock region (Figure~\ref{Post_AMR}) is more robust with respect to the spatial resolution, for a subtle numerical reason: the increase of $v_{\mathrm{rms}}$ (and of $P_{\mathrm{turb}}$) leads to an increase also in the numerical dissipation and thus to an increase in temperature and hydrodynamical pressure, roughly balancing the turbulent pressure in the pressure ratio (equation~\ref{support}). An analysis similar to Figure~\ref{Central_8AMR} has been performed using the norm of vorticity as diagnostic, with comparable results.

Finally, we observe that the previous resolution study can be interpreted as a verification on our simulations, namely it shows that our tool accurately represents the conceptual model (cf.~\citealt{Calder_ApjS_2002}). A further, important step would be the validation, i.e.~demonstrating that our setup appropriately describes the nature of turbulence in the ICM flow. Our study explicitly focuses on the generation of turbulence produced by merger shock, and to this aim the simulation setup was defined. On the other hand, it is known that other stirring mechanisms can inject turbulence in the ICM. How well can they be modeled within our setup, and how much do they contribute to the production of turbulence in merging clusters?

We believe that, by definition, the baroclinic vorticity production is the main stirrer in the post-shock region, and that our setup can follow it in a reliable way, as described above. By extending this analysis to the cluster core, other stirring mechanisms have to be considered. 

Turbulence production by AGN outflows \citep{Heinz_MNRAS_06,Sijacki_MNRAS_06a} has not been included in the presented setup, but it is not supposed to play a significant role in unrelaxed, non-cool core clusters. Minor mergers are also known to effectively inject turbulence in the ICM, but only with a small volume filling factor and in a rather localized way \citep{Subramanian_2006_MNRAS,Iapichino_2008_MNRAS,Maier_2009_ApJ}. These mergers inject turbulence via shearing instabilities in the wake of the accreted substructure, and this process is probably the most prone to be under resolved in hydrodynamical simulations.

\begin{figure*}
\epsscale{1}
\plotone{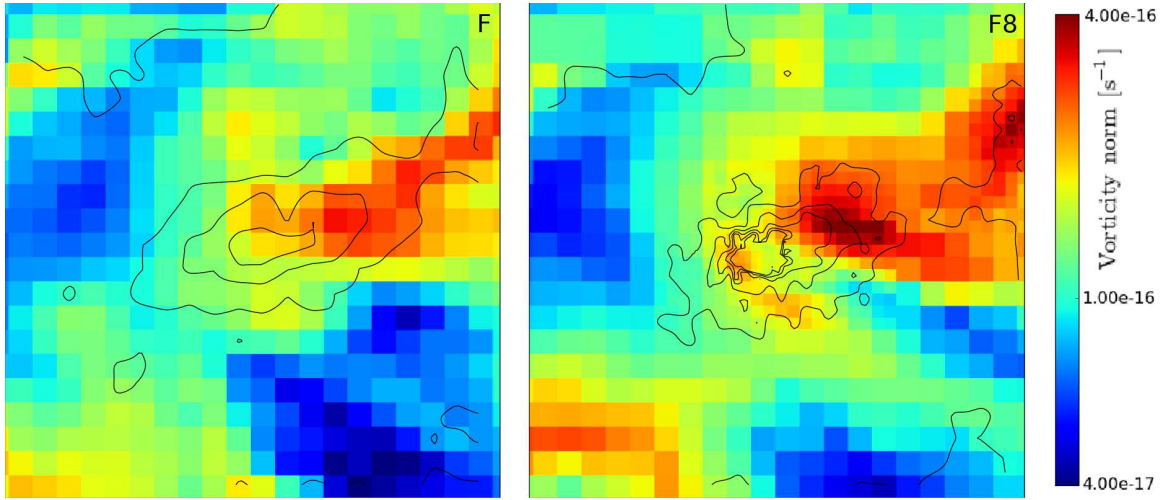}\\
\caption{Projections of the vorticity norm, with DM density contours overlayed, at $z = 0.1$, for the runs $F$ (left) and $F8$ (right). The projected volume has a size of (640 kpc $h^{-1})^3$. The volume is approximately centered at the location of maximum DM density, and contains the sphere for the analysis in the cluster core, as defined in
Section~\ref{scaling}.}
\label{vort-resol}
\end{figure*}

We checked therefore if the higher core resolution in run $F8$ would
result in a more accurate modeling of both the vorticity and the
turbulence generated during minor merger events and, in turn, to a
larger turbulent pressure support in the cluster core.  We find that,
despite the higher resolution of run $F8$ with respect to run $F$, the
evolution of both vorticity and the turbulence is qualitatively the
same in the two cases. To illustrate
the point in Figure~\ref{Central_8AMR} we compare the evolution of
$P_{\mathrm{turb}}/P_{\mathrm{tot}}$ for both run $F8$ and run $F$
(see caption for details). We only notice quantitative differences
which, at most, are at the level of $10$ to $20\%$.
In order to further investigate the resolution issues, in Figure~\ref{vort-resol} projections of the vorticity norm in the cluster core are compared, together with contours of DM density (derived from the N-body particles with the cloud-in-cell interpolation technique). The comparison is performed at $z=0.1$, corresponding to one of the largest discrepancies between runs $F$ and $F8$ in Figure~\ref{Central_8AMR}, at a lookback time of about $1.3\ \mathrm{Gyr}$. From Figure~\ref{vort-resol} we infer that the differences are mainly due to small scale structures which are better resolved in the $F8$ run, as shown clearly from the DM density contours. According to this result we conclude
that turbulence injected by the major merger is the leading
contribution in the core of this cluster. This class of mergers acts
mainly on length scales which are reliably resolved in our reference
setup. Other stirring mechanisms, although possibly underresolved or not resolved at all (e.g., turbulence produced in the wakes of cluster galaxies; \citealt{Bregman_ApJ_89,Kim_Apjl_07}) are not expected to play a crucial role in turbulence injection in cores of merging clusters.

\section{Discussion and conclusions}
\label{disc}

In this work the production of turbulence in cluster major mergers has been studied with hydrodynamical AMR simulations, with emphasis on turbulence injection in the post-shock region and in the cluster core. Representative mergers were selected in the simulated cosmological volume, and re-simulated in a setup specially designed for resolving the post-shock turbulence.

The morphological analysis was based on a reference computation (Section~\ref{morphology}) with typical features. Right after the two sub-clusters fall together, a merger shock is launched from the center of the newly formed structure, in a roughly symmetric fashion. The shock and the associated compression are the main agents of the conversion of gravitational to internal energy in the merging process. The values of the shock Mach number $\mathcal{M}$ are smaller than the expectations for external accretion shocks, but generally larger than the typical values for internal shocks (cf.~\citealt{Miniati2_ApJ_2000}).

The heating is particularly effective past the shock, in a region with a size of about $300\ \mathrm{kpc}\ h^{-1}$. In this region, as a byproduct of the merger event, some fraction of the energy is injected in the ICM as turbulent motions. As reported in Figure~\ref{press_ratio}, the energy contribution of turbulence in this region is at the level of few percent of $e_{\mathrm{int}}$ (at a length scale of $256\ \mathrm{kpc}\ h^{-1}$) for a timescale of $1.5\ \mathrm{Gyr}$.

As for the turbulence at the cluster core, at a length scale of $512\ \mathrm{kpc}\ h^{-1}$ the turbulent pressure support (as inferred by the velocity dispersion) is about $20 \%$ of the total pressure. The turbulent to total pressure ratio remains larger than $10 \%$ on a timescale of $2\ \mathrm{Gyr}$, and above the threshold of a nearly
relaxed cluster (about $5 \%$) for $4\ \mathrm{Gyr}$, until $z = 0$. Strictly related to this point is the relatively long timescale for virialization \citep{Poole_2006_MNRAS}, as deduced by the slow temperature increase noticeable in Figure~\ref{rlx_actv_turb}$c$.  The energy content of the turbulent motions in the ICM is comparable with previous investigations \citep{Dolag_MNRAS_2005,Vazza_2006_MNRAS,Iapichino_2008_MNRAS,Lau_ApJ_2010} and with values inferred from observations \citep{Schuecker_2004_A&A,Churazov_2004_MNRAS,Churazov_2008_MNRAS,Werner_MNRAS_2009}.

The analysis presented above shows that a major merger event
can affect the whole cluster evolution for about $4\ \mathrm{Gyr}$ or more, a significant fraction of the cluster history, and much longer than the  shock propagation timescale. A consequent and interesting question would pertain to any observable imprint that such event leaves on the cluster structure. A closely related issue concerns the bimodality in the correlation between radio and X-ray luminosity of clusters showing a radio halo, discussed by \citet{Brunetti_2007_ApJL} and \citet{Brunetti_2009_A&A}. As shown in these references, a complete sample of nearby
ROSAT flux-limited clusters with $L_X$ $(0.1\;-\;2.4 \; \mathrm{keV}) > 5 \times 10^{44} \; \mathrm{erg \; s^{-1}}$  apparently show bimodality, i.e. only a small fraction of these clusters have a detectable radio halo at 1.4 GHz and their radio and X-ray luminosities are correlated, while the rest of the sample have no detectable radio halo, and only upper limits. In the light of the turbulent re-acceleration theory, this bimodality was interpreted as originating from the short timescale (of the order of $1\ \mathrm{Gyr}$) of the turbulence driving and decay. It is currently unclear how this acceleration scenario and the observed bimodality could be reconciled with the theoretical evidences of long turbulence decay timescales, presented in our work. Addressing this issue requires a careful account of the drop in acceleration efficiency following the turbulence decay, which is beyond the scope of this paper.

A more comprehensive comparison with the radio observations would require the coupling of the used numerical scheme with some model of particle acceleration \citep{Miniati_2001_CPC,Miniati_2007_JCP}, but this is not addressed in this study. Moreover, our simulations setup has been especially designed for resolving the merger shocks, and therefore the spatial resolution in the cluster core (roughly corresponding with the effective resolution) is relatively low. We will improve upon this numerical shortcoming in future works, although the resolution study in Section~\ref{resol} shows convergence of our results.

The scaling study in Section~\ref{scaling} extends the works of \citet{Cassano_2005_MNRAS} and \citet{Vazza_2006_MNRAS} to focus on the special case of major merger: the results, though based on a small simulation sample, are consistent with the proportionality between internal and turbulence energy, since both scale with the cluster mass with a dependence close to $M^{5/3}$. This makes sense, because it is the same scaling law holding for the cluster thermal energy in the self-similar model. In other words, thermalization and injection of turbulence are two faces of a same physical mechanism, the formation of cosmological structure through accretion and mergers.

The results on the shock propagation are comparable with similar studies performed with idealized simulation setups \citep{Roettiger_1993_ApJL,Roettiger_1996_ApJ,Roettiger_1997_ApJS,Ricker_2001_ApJ}. The most relevant difference with those setups is the interaction between the outgoing shock and the filamentary structure which connects the newly formed cluster with the surrounding cosmic web. The shock propagation is thus hindered in the direction of the filaments, resulting in arc-like shapes in temperature and vorticity slices (Figures~\ref{temperature}~and~\ref{vorticity}).

\begin{figure*}
\epsscale{1}
\plotone{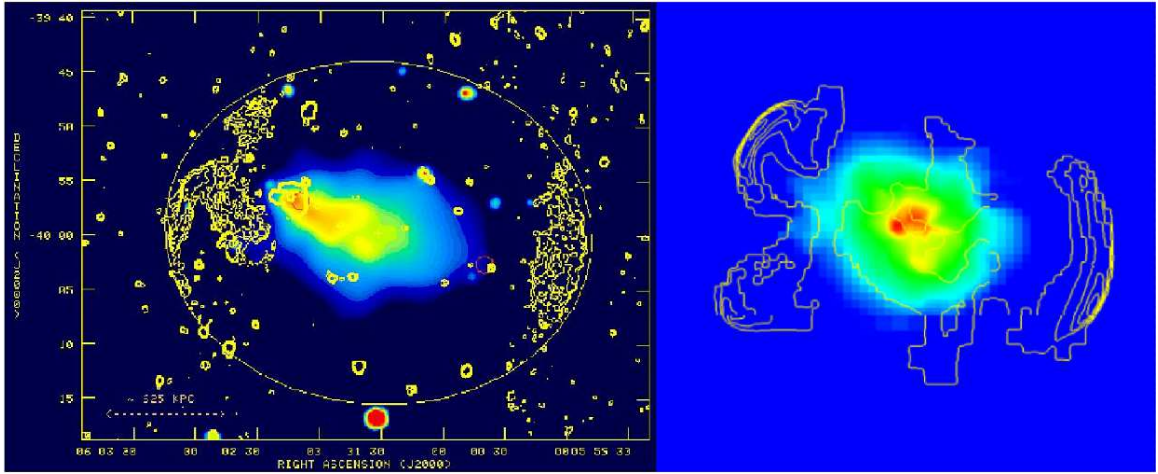}
\caption{Morphological comparison between the observed cluster A3376 (left) and the simulated cluster $F$ (right). Left panel: composite map of radio and X-ray emissions from the cluster A3376. The VLA observed 1.4 GHz total intensity contours (yellow) are: 0.12, 0.24, 0.48, and 1 mJy/beam. (beam: 20'' FWHM). The large ellipse shows an elliptical fit to the peripheral giant radio structure. The central color image depicts the thermal bremsstrahlung X-ray emission observed by the ROSAT PSPC instrument ($\sim$ 12 ks exposure, 0.14-2.0 keV band). The red circles mark the positions of the two brightest cluster galaxies. The image is taken from \citet{Bagchi_Sci_2006}, Reprinted with permission from AAAS. Right panel: simulated projection of cluster $F$ at $z = 0.05$, with a side of $7.68\ \mathrm{Mpc}\ h^{-1}$ and depth $2.56\ \mathrm{Mpc}\ h^{-1}$. The projected X-ray surface brightness (in the range $0.14 - 2.0\ \mathrm{keV}$) is drawn in colors, and the temperature is overlayed as contours.}
\label{morphology-fig}
\end{figure*}

Here we point out a striking morphological resemblance between
the shock wave structures obtained in our simulations and those actually
observed at radio wavelengths in a few well known merging clusters. In Figure~\ref{morphology-fig} the complex morphology of merging cluster A3376  (z = 0.046) observed with  VLA in radio and ROSAT in X-ray wavelength \citep{Bagchi_Sci_2006} is compared with the simulated merging cluster $F$, showing its projected temperature and X-ray emission maps. Our aim in this work was not of achieving a detailed match. Nevertheless, a Mpc-scale projected elliptical radio emission structure visible in A3376 is remarkably similar to the ellipsoidal merger shock
front morphology obtained in our simulation (Figure~\ref{morphology-fig}). Noticeably, 
the observed radio emission and the simulated shock fronts both  have double arc-like features, that are located at the cluster periphery with their concave sides facing the cluster center. Both the observed and the simulated cluster have a comparable mass of about $0.5 \times 10^{15}\ M_{\sun}$, major axis of elliptical structure of about $2-3 \ \mathrm{Mpc}\ h^{-1}$, virial radius of $\sim 1.4 \ \mathrm{Mpc}\ h^{-1}$, ICM temperature of about $ 5 \times 10^{7}\ \mathrm{K}$, length of the shock structures about $1\ \mathrm{Mpc}\ h^{-1}$ and, most noteworthy, both have an extension behind the shock front of about  $300\ \mathrm{kpc}\ h^{-1}$ \citep{Bagchi_Sci_2006}. The radio morphology matches the morphology of the simulated structure in temperature and, more importantly, in vorticity, which traces turbulence (see also Figures~\ref{temperature},~\ref{vorticity}~and~\ref{vorti_ext}).

An interesting feature in the eastern radio arc (left-hand side in Figure~\ref{morphology-fig}) of A3376 is a `notch-like' structure, where the arc apparently bent inwards towards the cluster center. It is natural, in the framework of the comparison with the simulations proposed above, to relate this observed morphology with the interaction of the merger shock with the filaments of the cosmic web, and the subsequent injection of turbulence at the interface between the post-shock region and the filament. A similar interpretation, obtained by the analysis of our simulation, can be applied also to another double radio arc with notch-like
features, in the merging cluster A3667 \citep{Rottgering_MNRAS_1997}.

Of course, a detailed comparison should make use of more sophisticated tools for converting the information from the hydrodynamical simulations in synthetic observations; this has not been performed here, but is left for future work.

The robustness of the presented results has been tested in Section~\ref{resol} against resolution effects. The implementation of additional numerical tools could increase the consistency of the turbulence modeling. Although grid codes are considered superior to SPH in this problem \citep{Agertz_2006_MNRAS,Mitchell_2009_MNRAS}, for both schemes it is not possible to spatially resolve the turbulent cascade down to the dissipation length scale. A way to consistently model the role of turbulence at sub-grid scales is provided by Large Eddy Simulations (LES). In these tools the unresolved scales are coupled to the resolved ones (where the hydrodynamical equations are solved) by means of a sub-grid scale model for turbulence. Examples of this technique in astrophysical problems are found in the simulations of the explosion of type Ia supernovae \citep{Niemeyer_1995_ApJ,Reinecke_2002_AAP,Schmidt_2006_A&A,Schmidt2_2006_A&A} as well as in other fields \citep{Pope_2008_MNRAS,Scannapieco_2008_ApJ}. Recently, \citet{Maier_2009_ApJ} coupled LES with AMR, developing a tool which is suitable for the study of turbulence generation in strongly clumped media such as galaxy cluster. The application of this novel technique to major merger simulations is potentially very interesting, since the turbulence energy content in these events is remarkable and needs to be properly accounted for. This problem will be addressed in a forthcoming work.

\acknowledgements
The computations described in this work were performed using the Enzo code, developed by the Laboratory for Computational Astrophysics at the University of California in San Diego (http://lca.ucsd.edu).
The numerical simulations were carried out on the SGI Altix 4700 {\it HLRB2} of the Leibniz Computing Center in Garching (Germany). S.P.~thanks the Deutsche Forschungsgemeinschaft (DFG) for providing adequate funding for the research and collaboration expenses, and the University of W\"urzburg Graduate Schools (UWGS) for providing a PhD finishing fellowship (STIBET Abschlu\ss stipendium). Thanks to G.~Brunetti for useful discussions, and to the anonymous referee for the valuable suggestions and constructive criticism which contributed to improve this work.

\bibliographystyle{apj}
\bibliography{ApJ_paper_bibtex}

\label{lastpage}

\end{document}